# Observation of coherent oxide precipitates in polycrystalline MgB$_2$

R. F. Klie, J. C. Idrobo and N. D. Browning
*Department of Physics (M/C 273), University of Illinois at Chicago, 845 West Taylor Street, Chicago, IL 60607-7059*

A.C. Serquis, Y. T. Zhu, X. Z. Liao and F. M. Mueller
*Superconductivity Technology Center, Los Alamos National Laboratory, Los Alamos, NM 87545*

**Abstract**

Here we describe the results of an atomic resolution study of oxygen incorporation into bulk MgB$_2$. We find that ~20-100 nm sized precipitates are formed by ordered substitution of oxygen atoms onto boron lattice sites, while the basic bulk MgB$_2$ crystal structure and orientation is preserved. The periodicity of the oxygen ordering is dictated by the oxygen concentration in the precipitates and primarily occurs in the (010) plane. The presence of these precipitates correlates well with an improved critical current density and superconducting transition behavior, implying that they act as pinning centers.

When MgB$_2$ was discovered in January 2001[1], its record-breaking transition temperature for a non-cuprate superconductor of 39K and remarkably high critical current sparked a large research effort to study the origin its the loss of conductivity[2], its transition behavior to the superconducting state[3], its critical current under various conditions[3], and the influence of the grain boundaries on the intergranular superconducting current[4]. According to initial findings, MgB$_2$ seems to resemble the properties of a conventional metallic superconductor, with remarkably high transition temperature, a coherence length of ~5 nm[5] and an absence of weak-links at the grain boundaries[4,6]. Furthermore, the presence of Mg non-stoichiometry and dopants can lead to different transition temperatures and critical currents[7,8]. Experiments performed on MgB$_2$ doped with different metals such as Mn, Co, Ni and Fe showed a significant decrease in T$_c$,[9]. However, the incorporation of oxygen into the structure appears to leave T$_c$ unaltered while increasing the critical current dramatically[10]. The incorporation of oxygen into the structure therefore represents a viable means of tailoring the superconducting properties of MgB$_2$.

In this letter, we present a direct atomic resolution[11] analysis of oxygen rich MgB$_2$, using analytical scanning transmission electron microscopy (STEM). In particular, we show the detailed structure of oxygen rich precipitates that were found in high concentrations within the bulk of MgB$_2$ grains. It was proposed earlier that an ordering of oxygen inside the bulk material is responsible for these defects that can act as pinning sides inside the grain, therefore causing a higher critical current and an improved supercon-ducting transition.[10] We find that oxygen replaces boron from the ideal MgB$_2$ structure, which causes a periodic superstructure to appear in both STEM and conventional TEM images. The combination of Z-contrast imaging[12] and EELS[13] allows us now to quantify the amount of oxygen present in the individual atomic columns and compare measured image contrast in the Z-contrast image with multi-slice image simulations[14] for the suggested MgBO structure.

The $MgB_2$ sample used for this study was prepared by solid-state reaction using amorphous B powder and Mg turnings in the atomic ratio 1:1. The B-powder was pressed into small pellets under 500 MPa and all the materials were wrapped in Ta foil. The sample was heated for 1 h at 900° C in a tube furnace under ultra high purity Ar. After an additional heating cycle from 500° C to 900° C (for more information see [7]) the material was slowly cooled to room temperature. The resulting poly-crystalline $MgB_2$ material contained grains with an average diameter approximately 1-4 μm.

The experimental Z-contrast images[12] and EEL spectra[13] in this study were obtained from a JEOL 2010F field emission STEM/TEM operating at 200kV.[15] The lens conditions in the microscope for imaging were defined for a probe-size of 0.14 nm (0.2 nm for spectroscopy), with a convergence angle of 13 mrad and a collection angle of 52 mrad. In this experimental condition, the incoherent high-angle annular dark field or "Z-contrast" image[12] allows the structure of the grains to be directly observed and the image can also be used to position the electron probe for EELS[11]. Core loss EELS probes the unoccupied density of states near the conduction band minimum and is exactly analogous to near edge X-ray absorption spectroscopy (XAS), but with a much higher spatial resolution afforded by an electron microscope. In the experiment performed here, the EEL spectra are acquired directly from the grain with characteristic core losses used to determine the composition. The samples were prepared for microscopy by the standard TEM specimen techniques of mechanical grinding (using the wedge technique) followed by ion-milling with 3 kV argon ions until electron transparency.

Figure 1(a) shows a low-magnification image of a typical precipitate observed in these samples. Figure 1(b) shows a high-resolution Z-contrast image taken from the bulk of $MgB_2$ grain close to the precipitate that was oriented along the [010] orientation (We will be using the three-index notation system to index the hexagonal $MgB_2$ crystal in this letter). The bright spots in this image that form a rectangular shape represent the pure Mg atomic columns. The B-columns are not visible in this image, due to the low scattering amplitude of B (intensity ratio with Mg of 0.17:1). EELS-spectra taken from this area (Figure 2) show the B core-loss peak with a fine structure typical for pure $MgB_2$ [6] with no oxygen signal being detected. This confirms that the bulk of the grains are composed of pure $MgB_2$ (within the ~5% detection limits of this technique). Figure 1(c) shows a Z-contrast image from the ~20 nm diameter precipitate that appeared brighter in low-magnification Z-contrast image shown in Figure 1 (a). Here again, the bright spots indicate the pure Mg-columns forming the rectangular frame of the $MgB_2$ unit cell. It can be clearly seen from this micrograph, that the spacing between the horizontal Mg columns is not constant throughout the image. Moreover, in every second column the spacing seems to be decreased, which corresponds to an increase in the background intensity between the two Mg columns. This phenomenon was observed in many other areas of the sample with varying number (n= 1,2,3...) of regular $MgB_2$ unit cells between the brighter B-columns.

Atomic resolution EEL spectra were taken from the bright and from the dark columns of the area shown in Figure 1 (c). Figure 2(a) shows the resulting B-K edge spectra from the precipitate. Each atomic resolution spectrum is a sum of 7 individual spectra with an acquisition time of 1 s and an energy resolution of 1.0 eV. The spectra are summed up to increase the signal-to-noise ratio, background subtracted and corrected for multiple-scattering contributions.[12] The B K-edge onset for all three spectra was determined to be at (187.2±0.5) eV. The B K-edge of the dark columns shows a fine-structure very similar to the bulk spectrum. The near edge fine structure of the bright column spectra shows clear differences to the bulk and the dark column spectra. The pre-peak intensity (labeled a) in the bright column is significantly higher and the splitting between the peaks (labeled b and c) is more obvious in this spectrum than it is in either the dark column or the bulk spectra. In addition, the shape of the main-peak appears broader.

Figure 2 (b) shows the oxygen core-loss spectrum taken from the same position. The spectra shown here represent a sum of 7 individual atomic resolution spectra, which are subsequently background subtracted and corrected for multiple scattering effects. The integrated K-edge intensities of the bright and dark columns reveal a (47.6±4.8) % higher oxygen concentration in the bright columns than in the dark columns. Unfortunately, the signal-to-noise ratio of the atomic resolution spectra is still very low, due to the large specimen thickness at the location of the precipitates. Hence, an interpretation of the differences in the near-edge fine-structure is not possible. Nevertheless, the increased intensity in the bright B-columns strongly suggests that a substantial amount of boron atoms in these atomic columns is replaced by oxygen. This effect would give rise to the observed change in contrast in the Z-contrast image and also describe the measured EEL spectra. The residual oxygen intensity in the dark columns can be explained by the fact that the tails of the electron probe expand into the adjacent unit cells and pick up the oxygen contribution from the bright columns.

Next, we will use multi-slice Z-contrast image simulations to verify the changes in image contrast caused by the oxygen ordering. The Z-contrast image simulations were performed using the program developed by Kirkland[13]. To generate the images, it is assumed that the beam in the microscope interacts with an effective potential of the specimen as a whole. Here, the effective potential of $MgB_2$ is assumed to be a linear superposition of the potentials for each atom, and this is calculated with a relativistic form of the Hartree-Fock[13] method. The algorithm creates a focused probe by taking into consideration the lens aberrations and the objective aperture. The electron probe then passes through the material and is modulated by the transmission function. The Fourier Transform of the transmitted wave function is then projected onto the annular dark field (ADF) detector and the square modulus of this wave function in the ADF detector is integrated over the entire scattered angle. Noise is then added to the resulting image to obtain a final simulated image that can be compared with the experiment.

The simulated images were obtained using conditions very similar to the experimental setup, where the acceleration voltage is $V_a$=200 kV, the convergence angle $\alpha_c$ ~15 mrad, and the spatial resolution is ~0.14nm. The high angle annular dark field detector range was chosen to be 52-200 mrad. The structure that was used for image simulation was derived from the ideal $MgB_2$ structure, i.e. not from the experimental images. Figure 3 (a) shows the simulated image of $MgB_2$ in the [010] orientation with the bright Mg atomic columns and the pure B intensities not distinguishable from the background noise level. For the oxygen ordered phase, we simply replaced one B atom in every second atomic column by oxygen, therefore obtaining an overall oxygen concentration of 54%. The resulting image is shown in figure 3 (b); arrows indicate the mixed B-O columns, where the oxygen columns can be recognized. The similarities between the experimental and the simulated images further confirm the proposed ordering of oxygen in the B columns.

In conclusion, the combination of atomic resolution Z-contrast imaging, EELS and theoretical image simulations unambiguously show that precipitates containing a coherent oxide phase of the form $MgBO-nMgB_2-MgBO$ (n=1,2…) occur in bulk $MgB_2$. In this letter, we concentrated on the n=2 phase, but various other phases (n=3-5) were observed that showed similar behavior. The periodicity of the mixed B-O columns in the precipitates is solely determined by the oxygen concentration (i.e. higher oxygen concentration, higher periodicity). In addition, several precipitates did not exhibit any change in contrast in the B atomic columns; the overall contrast was higher than in the bulk but homogeneous. The measured oxygen content in these phases suggested that these precipitates could contain ordered oxygen phases where the ordering orientation is not orthogonal to the electron beam direction.

From the results reported here, it seems conceivable that the oxygen substitution occurs in the bulk of the $MgB_2$ grains, where coherent ordered MBO precipitates are formed. These precipitates act as magnetic pinning centers, therefore increasing the critical current both at low and at high magnetization. The sharper decrease in the superconducting transition in this material is also caused by the high density of oxide precipitates in the bulk, in addition to the larger grain size and the absence of amorphous B-O phases at grain boundaries. Hence, it appears crucial for improving the superconducting behavior of future $MgB_2$ materials to control the formation coherent oxide precipitates inside the $MgB_2$ bulk material. Although $MgB_2$ does not show any weak link behavior, secondary phases formed at the grain boundaries might. Increasing the grain connectivity and removing secondary interfacial phases might further decrease the overall critical current and will be subject to future investigations.

This work is supported by the US Department of Energy under grant number DOE FG02 96ER45610. The experimental results were obtained on the JEOL 2010F operated by the Research Resources Center at UIC and funded by NSF.

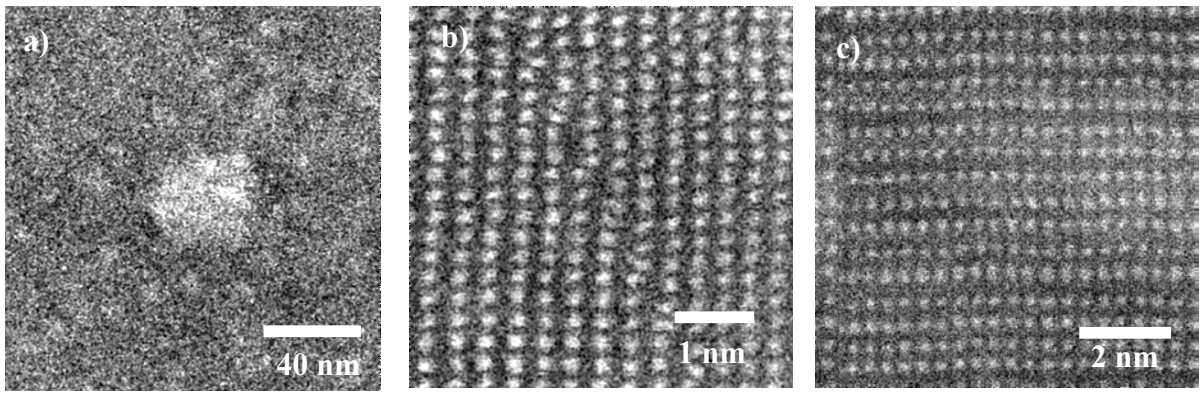

**Figure 1**: a) Low magnification Z-contrast micrograph showing the brighter precipitate; b) Z-contrast image of bulk $MgB_2$ in the [010] direction. The bright spots represent the Mg columns; pure B columns are not visible. c) Z-contrast image of the coherent oxide precipitates in the bulk of the $MgB_2$ [010]. A contrast variation in every second columns is visible.

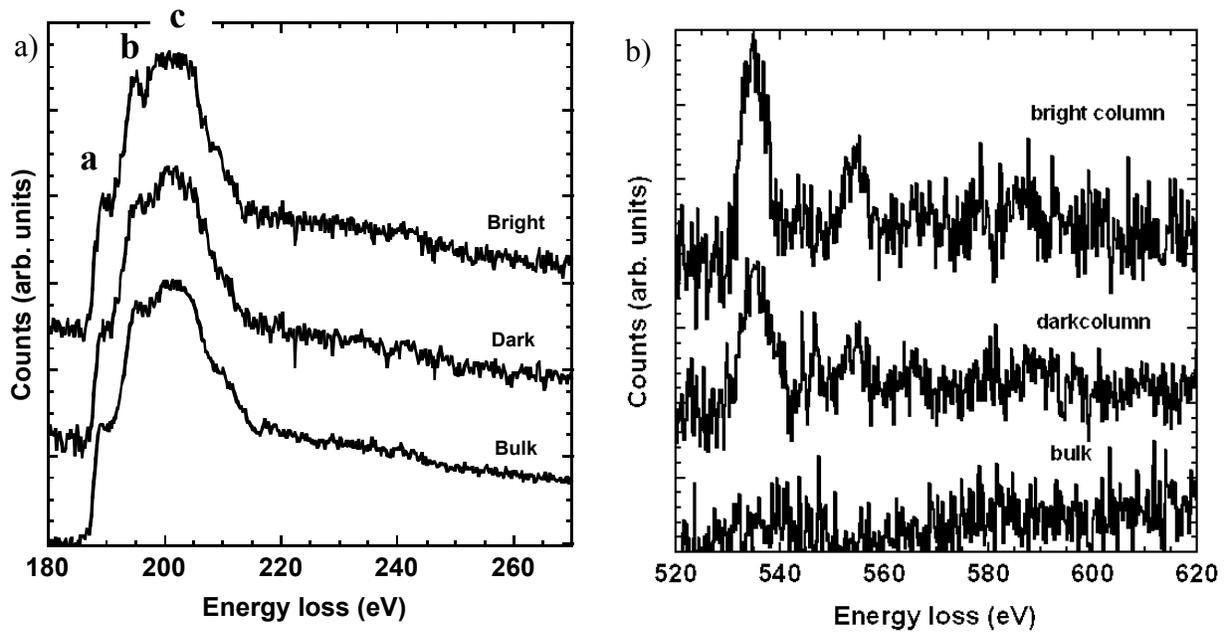

**Figure 2**: a) B-K edges from the bulk, and the bright and dark column in the oxide precipitates. b) O-K edges from the bright and dark columns of the oxide precipitates. Each spectrum is a sum of 7 spectra with an acquisition time of 2 s and an energy resolution of 1.0 eV per pixel.

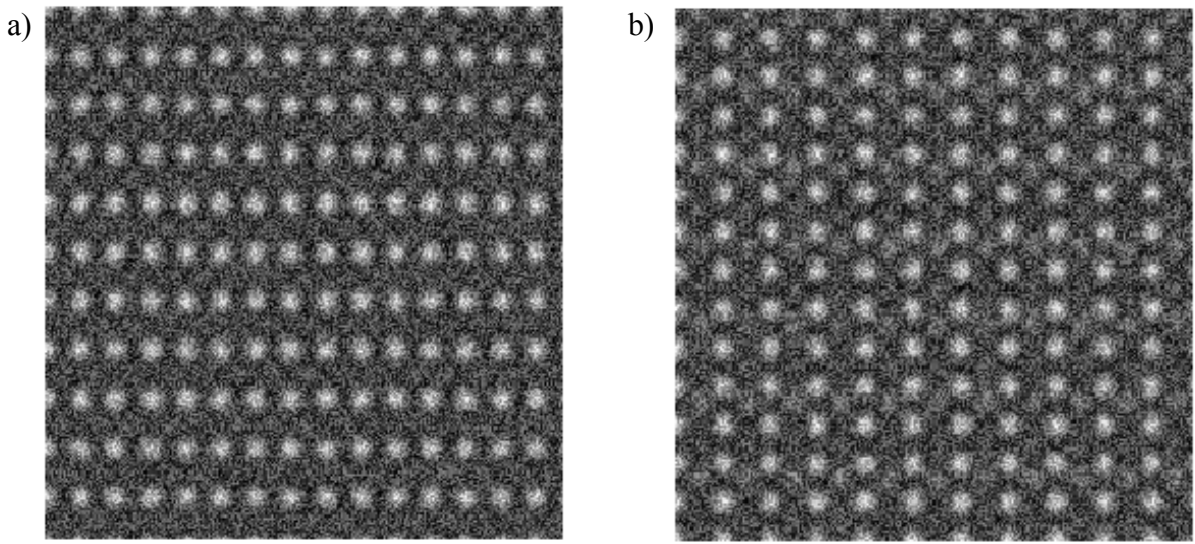

**Figure 3**: a) Simulated Z-contrast image from bulk $MgB_2$ in the [010] direction. b) Simulated Z-contrast image from the MgB2 bulk matrix with 54% oxygen substitution.